\documentclass[12pt]{article}

\usepackage[english]{babel}
\usepackage[utf8x]{inputenc}
\usepackage{amsmath}
\usepackage{graphicx}
\usepackage{hyperref}

\title{Evolution of choices over time:
The U.S. Presidential election 2012 and the NY City Mayoral Election, 2013}
\author{Mukkai Krishnamoorthy, Wesley Miller \\ Rensselaer Polytechnic Institute, Troy, NY \and Raju Krishnamoorthy \\ Columbia University, New York, NY}

\begin{document}
\maketitle

\begin{abstract}
We conducted surveys before and after the 2012 U.S. Presidential election and prior to the NY City Mayoral election in 2013. The surveys were done using Amazon Turk. This poster describes the results of our analysis of the surveys and predicts the winner of the NY City Mayoral Election. 

\end{abstract}

\section{Introduction}
Several authors have published analyses of poll and survey data of the U.S. Presidential election 2012 \cite{silver}\cite{morris}, with vastly different predictions. Both sampling differences in voter populations and differences in methods of analysis may have accounted for the difference in the prediction. In the current experiment, we perform a survey and an indirect analysis, using Amazon Turk, as opposed to analysis of direct voter preferences using Amazon Turk \cite{cuff}\cite{sides}. We have focused on voters'  two most pressing concerns, one national and the other international, from a choice of 5 in each category.

A lot of data analysis of the presidential election in 2012 has been done. In
particular, please see \cite{silver}. Other data analysis \cite{morris} did not fare that well. One of
the reasons for such a difference could be the sampling of voter population.
Another reason could be due to data analysis. Please see these two other
papers for using Amazon Turk \cite{cuff}\cite{sides} in Election
Prediction. As opposed tp these two papers, we do an indirect analysis - not
measuring direct voter preferences.The perspective voters were asked
to give on a pressing national and international concern. Our approach is to do a
seconday analysis, concentrating more on the most important issue (one of a
national concern and the other of international concern among choices
of 5 in each category) that the surveyed people had in their minds.

These are the limitations on our data:
\begin{itemize}
\item Did all those surveyed for the presidential election 2012, vote?
\item Were the same people surveyed before and after the election? 
\item Will those surveyed for the NY City Mayoral Primary vote?
\item Sample size is extremely small, compared to the voting population (200 for the US Presidential election and 100 for the NY City Mayoral)
\end{itemize}

Even with these limitations, we feel that it exhibits interesting patterns.

\section{Data Collection}
The data was collected through Amazon Mechanical Turk. Our pool of participants was people of voting age in the United States, the voting population. We selected 200 people before the election and asked each one of them which domestic issue would affect his/her voting decision most when considering the economy, healthcare, tax reform, education, and national security. We then asked each one of them which international issue would have the greatest effect on his/her voting decision when considering withdrawal from wars, international security, global trade issues, more active engagement, and international partners. Once the election was over and President Obama had been re-elected, we posted the same survey for 200 more people, asking this time which of the issues affected their decisions the most when they voted. The third set of data was collected from a pool of 100 participants and asked the same set of questions (that we asked during the presidential election).

\section{ Amazon Turk Experimental Results}
\begin{tabular}{ l }
Presidential Before \\
\begin{tabular}{ l  c }
   National & International \\
   \begin{tabular}{ l | c }
     \hline
     The Economy & 145 \\ \hline
     Healthcare & 30 \\ \hline
     Education & 11 \\ \hline
     Tax Reform & 12 \\ \hline
     National Security & 2 \\
     \hline
   \end{tabular}
   & 
   \begin{tabular}{ l | c }
     \hline
     Withdrawal from Wars & 109 \\ \hline
     International Security & 39 \\ \hline
     Global Trade Issues & 40 \\ \hline
     More Active Engagement & 6 \\ \hline
     International Partners & 6 \\
     \hline
   \end{tabular}
 \end{tabular}\\\\\hline
 Presidential After \\
 \begin{tabular}{ l  c }
   National & International \\
   \begin{tabular}{ l | c }
     \hline
     The Economy & 147 \\ \hline
     Healthcare & 36 \\ \hline
     Education & 9 \\ \hline
     Tax Reform & 7 \\ \hline
     National Security & 1 \\
     \hline
   \end{tabular}
   & 
   \begin{tabular}{ l | c }
     \hline
     Withdrawal from Wars & 105 \\ \hline
     International Security & 43 \\ \hline
     Global Trade Issues & 42 \\ \hline
     More Active Engagement & 3 \\ \hline
     International Partners & 7 \\
     \hline
   \end{tabular}
 \end{tabular}\\\\\hline
 Mayoral Before \\
 \begin{tabular}{ l  c }
   City & National \\
   \begin{tabular}{ l | c }
     \hline
     The Economy & 71 \\ \hline
     Healthcare & 15 \\ \hline
     Education & 7 \\ \hline
     Tax Reform & 7 \\ \hline
     City Security & 0 \\
     \hline
   \end{tabular}
   & 
   \begin{tabular}{ l | c }
     \hline
     Withdrawal from Wars & 39 \\ \hline
     National Security & 29 \\ \hline
     Global Trade Issues & 15 \\ \hline
     More Active Engagement & 11 \\ \hline
     State Partners & 6 \\
     \hline
   \end{tabular}
 \end{tabular}
\end{tabular}

\section{Observations}
 The results of ordering all the surveys (shown in the previous section) remain the same. This tells us about the general anxiety level of the people. Even aftera gap of one year, rankings (of concerns) did not change.  Even though our sample size is small, we believe in the authenticity of data because the survey after the election did not vary even slightly from those taken before the election. What is more striking is the approximate percentage of people prefering the choices (in almost all three cases). There is a small change of  percentages of healthcare and International Security (before and after election) as both are hot button issues among the voting population.
 
 Based on presidential outcome, we conjecture that Mr. de Blasio, the democratic party'schoice, will win the November Mayoral election in New York City (Nate Silver predicted Ms. Quinn, who lost in the democratic primaries \cite{natesnyc}).

\section{Conclusion}
There was evident variation in the sets of data; however, it is so small that it can be attributed to the small sample size relative to the voting population. There is no evidence of a change in voter opinion between the two surveys and the mayoral election.

\section{Acknowledgements}

The Authors wish to thank Dr Janaki Krishnamoorthy and Prof Gurpur Prabhu with their constructive suggestions and editing the manuscript. First author wishes to thank Mr. Sean O'Sullivan for establishing Rensselaer Center for Open Source Software where part of this work is carried out.

\end{document}